\documentclass[pre,floatfix,twocolumn,showpacs]{revtex4}

\usepackage{amsmath}
\usepackage{amssymb}
\usepackage{graphicx}
\usepackage{graphicx,psfrag}

\usepackage{verbatim}

\newcommand{\ie}{\emph{i.}$\,$\emph{e.}}
\newcommand{\eg}{\emph{e.}$\,$\emph{g.}}
\newcommand{\etal}{\emph{et}$\,$\emph{al.}}

\newcommand{\muem}{$\mu\text{m}\,$}
\newcommand{\kT}{k_{\text{B}}T}

\newcommand{\romd}{{\operatorname{d}}}

\newcommand{\VECe}{{\boldsymbol{e}}}
\newcommand{\VECf}{{\boldsymbol{f}}}
\newcommand{\VECl}{{\boldsymbol{l}}}
\newcommand{\VECn}{{\boldsymbol{n}}}

\newcommand{\VECnab}{{\boldsymbol{\nabla}}}

\newcommand{\At}{A_{\text{contact}}}

\newcommand{\Rt}{R_{\text{tip}}}
\newcommand{\Rp}{R_{\text{pore}}}
\newcommand{\tildeRp}{\tilde{R}_{\text{pore}}}
\newcommand{\Rint}{R_{\text{int}}}
\newcommand{\wthr}{\tilde{w}_{\text{thr}}}

\newcommand{\CALK}{\mathcal{K}}

\begin{document}

\title{How to determine local elastic properties of lipid bilayer membranes from 
atomic-force-microscope measurements: A theoretical analysis}

\author{Davood Norouzi}
\affiliation{Institute for Advanced Studies in Basic Sciences, %
             P.O.Box: 45195-1159, %
             Zanjan, %
             Iran}
\author{Martin Michael M\"uller}
\author{Markus Deserno}
\affiliation{Max-Planck-Institut f\"ur Polymerforschung, %
             Ackermannweg 10, %
             55128 Mainz, %
             Germany}

\date{\today}
\begin{abstract}
  Measurements with an atomic force microscope (AFM) offer a direct way to 
  probe elastic properties of lipid bilayer membranes locally: provided 
  the underlying stress-strain relation is known, material parameters such as 
  surface tension or bending rigidity may be deduced. 
  In a recent experiment a pore-spanning membrane was poked with an AFM tip, 
  yielding a linear behavior of the force-indentation curves. A theoretical  
  model for this case is presented here which describes these curves in the 
  framework of Helfrich theory. The linear behavior of the measurements is 
  reproduced if one neglects the influence of adhesion between tip and membrane. 
  Including it via an adhesion balance changes the situation significantly: 
  force-distance curves cease to be linear, hysteresis and nonzero detachment 
  forces can show up. The characteristics of this rich scenario are discussed 
  in detail in this article.
\end{abstract}


\maketitle


\section{Introduction\label{sec:introduction}}

Lipid bilayer membranes constitute one of the most fundamental
components of all living cells. Apart from their obvious
\emph{structural} role in organizing distinct biochemical
compartments, their contributions to essential \emph{functions}
such as protein organization, sorting, or signalling are now well
documented \cite{Lodish}.  In fact, their tasks significantly
exceed mere passive separation or solubilization of proteins,
since often \emph{mechanical} membrane properties are intricately
linked to these biological functions, most visibly in all cases
which go along with distinct membrane deformations, such as exo-
and endocytosis \cite{Marsh01}, vesiculation
\cite{Kirchhausen93,Robinson97,McMahonGallop05}, viral budding
\cite{GaHe98}, cytoskeleton interaction \cite{LuHi92}, and
cytokinesis \cite{UmEm99}. Consequently, a quantitative knowledge
of the material parameters which characterize a membrane's elastic
response -- most notably the bending modulus $\kappa$ -- is also
biologically desirable.

Several methods for the experimental determination of $\kappa$
have been proposed, such as monitoring the spectrum of thermal
undulations via light microscopy \cite{BroLen75,FaMiMeBiBo89},
analyzing the relative area change of vesicles under micropipette
aspiration \cite{EvaRaw90,RaOlMcNeEv00}, or measuring the force
required to pull thin membrane tethers
\cite{DaiShe95_99,HoShDaSh96,CuDeBaNa05,MaeSenMeg02}.  With the 
possible exception of the tether experiments, these techniques are
\emph{global} in nature, \ie, they supply information averaged
over millions of lipids, if not over entire vesicles or cells.
Yet, in a biological context this may be insufficient
\cite{HeiWir04}. For instance, membrane properties such as their
lipid composition or bilayer phase (and thus mechanical rigidity)
have been proposed to vary on submicroscopic length scales
\cite{BroLon97_98,SiIk97}. Despite being biologically enticing,
this suggestion, known as the ``raft hypothesis'', has repeatedly
come under critical scrutiny \cite{Munro03,Nichols05,Hancock06},
precisely because the existence of such small domains is extremely
hard to prove.

An obvious tool to obtain mechanical information for small samples
is the atomic force microscope (AFM) \cite{BuCaKa05}, and it has
indeed been used to probe cell elastic properties (such as for
instance their Young modulus) \cite{Radmacher02,Costa03}. Yet,
obtaining truly \emph{local} information still poses a formidable
challenge. Apart from several complications associated with the
inhomogeneous cell surface and intra-cellular structures beneath
the lipid bilayer, one particularly notable difficulty is that the
basically unknown boundary conditions of the cell membrane away
from the spot where the AFM tip indents it preclude a quantitative
interpretation of the measured force, \ie\ a clean way to
translate this force into (local) material properties.  To
overcome this problem Steltenkamp \etal\ have recently suggested
to spread the cell membrane over an adhesive substrate which
features circular pores of well-defined radius
\cite{Steltenkamp_etal06}. Poking the resulting ``nanodrums''
would then constitute an elasto-mechanical experiment with
precisely defined geometry. Using simple model membranes, the
authors could in fact show that a quantitative description of such
measurements is possible using the standard continuum
curvature-elastic membrane model due to Canham \cite{Canham70} and
Helfrich \cite{Helfrich73}.

Spreading a cellular membrane without erasing interesting local
lipid structures obviously poses an experimental challenge; but
the setup also faces another problem which has its origin in an
``elastic curiosity'':  even significant indentations, which
require the full \emph{nonlinear} version of the Helfrich shape
equations for their correct description, end up displaying
force-distance-curves which are more or less \emph{linear}---a
finding in accord with the initial regime of membrane tether
pulling \cite{Goldstein02,Derenyi02}. Yet, this simple functional 
form makes a unique extraction of the two main mechanical 
properties, tension and bending modulus, difficult.  Is the 
nanodrum setup thus futile?

In the present work we develop the theoretical basis for a slight
extension of the nanodrum experiment that will help to overcome
this impasse.  We will show that an additional \emph{adhesion}
between the AFM tip and the pore-spanning membrane will change the
situation very significantly---quantitatively and qualitatively.
Force-distance-curves cease to be linear, hysteresis, nonzero
detachment forces and membrane overhangs can show up, and various
new stable and unstable equilibrium branches emerge.  The
magnitude and characteristics of all these new effects can be
quantitatively predicted using well established techniques which
have previously been used successfully to study vesicle shapes
\cite{Svetina89,Miao91,Berndl91,Hamiltonian95,Seifert97}, vesicle
adhesion \cite{Seifert90,Seifert95}, colloidal wrapping
\cite{wrapping1,wrapping2,wrapping3} or tether pulling
\cite{Goldstein02,Derenyi02,Smith03,Smith04,Koster05}.

\begin{figure}
\includegraphics[scale=1]{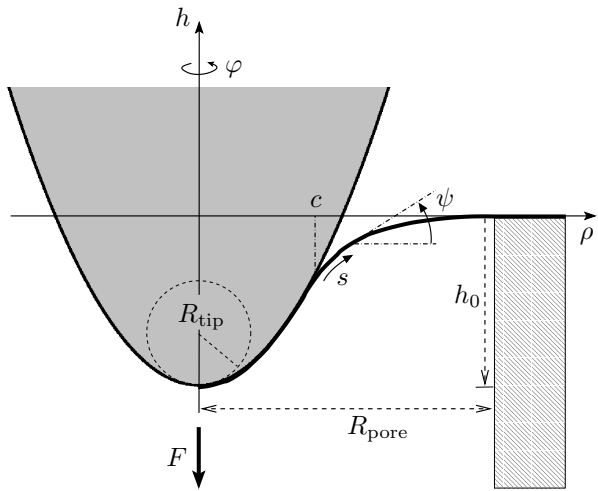}
  \caption{Illustration of the geometry. A parabolic tip with curvature radius
  $\Rt$ indents a pore-spanning membrane with a force $F$ to a certain
  depth $h_{0}$. The radius of the pore is $\Rp$. The membrane detaches from the tip
  at a radial distance $\rho=c$. The two possible parametrizations
  $h(\rho)$ and $\psi(s)$ are explained in the beginning of
  Chapter~\ref{sec:shapeeqn}.}
  \label{fig:poregeometry}
\end{figure}

The key ``ingredient'' underlying most of the new physics is the
fact that the membrane can \emph{choose} its point of detachment
from the AFM tip.  Unlike in the existing point force descriptions
\cite{Goldstein02,Derenyi02}, in which a certain (pushing or
pulling) force is applied at one point of the membrane, our
description accounts for the fact that the generally nonvanishing
interaction energy per unit area between tip and membrane
co-determines their contact area over which they transmit forces,
and thereby influence the entire force-distance-curve. What may at
first seem like a minor modification of boundary conditions
quickly proves to open a very rich and partly also complicated
scenario, whose various facets may subsequently be used to extract
information about the membrane.  In fact, Smith \etal\
\cite{Smith03,Smith04} have demonstrated in a related situation
that the competition between adhesion and tether pulling for
substrate-bound vesicles gives rise to various first- and
second-order transitions, details of which depend in a predictable
way on the experimental setup.  In our case we will for instance
find snap-on and snap-off events between tip and membrane, which
rest on the fact that binding is \emph{not} pre-determined, and
whose correct description is very important for reliably
interpreting any AFM force experiments. Moreover, we will also see
that the very occurrence of tethers is a much more subtle
phenomenon, since an adhering membrane pulled upwards may in fact
prefer to \emph{detach} rather than being pulled into a tether---a
question treated previously (and on the linear level) by
Boulbitch~\cite{Boulbitch}.

Our paper is organized as follows: in Chapter~\ref{sec:themodel}
we introduce the model of our system and discuss the relevant
energies. In Chapter~\ref{sec:shapeeqn} we present the equations
that have to be solved in order to find membrane profiles,
force-indentation curves and detachment forces.  This will also include 
a treatment of the nonlinear case which was only mentioned very briefly in
Ref.~\cite{Steltenkamp_etal06}. In Chapter~\ref{sec:results} the
results of our calculations are summarized and compared to
existing \cite{Steltenkamp_etal06} measurements. We end in 
Chapter~\ref{sec:discussion} with a discussion how the predictions 
for indentation and adhesion characteristics can be
used to extract material properties in future experiments.


\section{The model\label{sec:themodel}}

\subsection{Geometry of the system}

We consider a flat solid substrate with a circular pore of radius
$\Rp$. A lipid bilayer membrane is adsorbed onto the substrate and
spans the pore. In the situation we want to analyze an AFM tip is
used to probe the properties of the free pore-spanning membrane.
We assume that the tip has a parabolic shape with curvature radius
$\Rt$ at its apex. Furthermore, we restrict ourselves to the
static axisymmetric situation in which the tip pokes the
free-standing membrane exactly in the middle of the pore (see
Fig.~\ref{fig:poregeometry}).

For a certain downward force $F>0$ the membrane is indented to a
corresponding depth $h_{0}>0$ which is measured from the plane of
the substrate to the depth of the apex of the tip. Note that it is
also possible to pull the membrane up with a force $F<0$ in the
opposite direction if attractive interactions attach the membrane
to the tip.

In the following, we will model the bilayer as a two-dimensional
surface. This is a valid approach provided the thickness of the
membrane (approx. 5 nm) is much smaller than (i) the membrane's
lateral extension as well as (ii) length scales of interest such
as local radii of curvature.

With this geometric setup in mind, let us now consider the
different energy contributions we want to include in our model.


\subsection{Energy considerations\label{subsec:energyparts}}

The total energy of the system ``pore-tip'' comprises different
contributions: the membrane is under a \emph{lateral tension}
$\sigma$. To pull excess area into the pore, work has to be done
against the adhesion between membrane and flat substrate
\cite{substrate}. It is given by $\sigma$ times the excess area
\cite{lateraltension}. Additionally, a \emph{curvature energy} is
associated with the membrane. According to Canham \cite{Canham70}
and Helfrich \cite{Helfrich73} the Hamiltonian for an up-down
symmetric membrane is then
\begin{equation}
  E_{\text{elast}}
    = \int_{\Sigma} \!\!\romd A \; \big(\sigma + \frac{\kappa}{2} K^{2}
      + \bar{\kappa} K_{\text{G}} \big)
  \; ,
  \label{eq:Helfrich}
\end{equation}
where $\Sigma$ denotes the surface of the membrane part which
spans the pore. The proportionality constants $\kappa$ and
$\bar{\kappa}$ are called bending rigidity and saddle-splay
modulus, respectively. The Gaussian curvature $K_{\text{G}}$ is
the product of the two principal curvatures whereas $K$ is their
sum \cite{DifferentialGeometry,signconvention}. Note that the last
term of energy~(\ref{eq:Helfrich}) yields zero in our specific
problem \cite{Gaussiancurvature}.

With the help of the two material constants $\sigma$ and $\kappa$
one can define a characteristic lengthscale
\begin{equation}
  \lambda := \sqrt{\frac{\kappa}{\sigma}} \; ,
  \label{eq:characteristiclengthscale}
\end{equation}
which does not depend on geometric boundary conditions such as the
radius of the tip or the pore but only on properties of the
membrane. On scales larger than $\lambda$ tension is the more
important energy contributions; on smaller scales bending
dominates.

Apart from tension and bending, an \emph{adhesion between tip and
membrane} may contribute to the total energy. We assume that it is
proportional to the contact area $\At$ between tip and membrane
with a proportionality constant $w$, the adhesion energy per area.

If the indentation $h_{0}$ is given and one wants to determine the 
force $F$, the total energy can thus be written as
\begin{equation}
  E^{h_{0}}_{\text{total}} = \int_{\Sigma} \!\! \romd A \;
    \big(\sigma + \frac{\kappa}{2} K^{2}\big)
      - w \At
  \ .
  \label{eq:Etotal}
\end{equation}
Under certain circumstances, however, it is more convenient to
consider the problem for a given force $F$. Both ensembles
(``constant indentation'' \emph{vs.} ``constant force'') are
connected via a Legendre transformation \cite{Smith03,Smith04},
$E^{F}_{\text{total}} = E^{h_0}_{\text{total}} - F h_{0}$. While
the ground states one obtains for the two ensembles will be the
same, questions of stability depend on the ensemble: a profile
found to be stable under constant height conditions is not
necessarily stable under constant force conditions.

The route we want to follow here in order to find
force-indentation curves is to determine the equilibrium shapes of
the non-bound section of the membrane via a functional
minimization. The energy contributions caused by the bounded
section of the membrane enter via the appropriate boundary
conditions (see Chapter~\ref{sec:shapeeqn} and
Appendix~\ref{app:boundaryconditions}). These imply that the
contact point $c$ is not known a priori but has to be determined
as well (``moving boundary problem''). In the next section we will
show how one can set up the appropriate mathematical formulation
of the problem to get membrane profiles and force-indentation
curves.


\section{Shape equation and appropriate boundary conditions\label{sec:shapeeqn}}

To describe the shape of the membrane we use two different kinds of
parametrization (see Fig.~\ref{fig:poregeometry}): for the linear
approximation it is sufficient to use ``Monge'' gauge where the
position of the membrane is given by a height $h(\rho)$ above (or
below) the underlying reference plane. The disadvantage of this
parametrization is that it does not allow for ``overhangs''. Since
these may be present in the full nonlinear problem, we will use
the ``angle-arclength'' parametrization in the exact calculations:
the angle $\psi(s)$ with respect to the horizontal substrate as a
function of arclength $s$ fully describes the shape.

\subsection{Linear approximations\label{subsec:linearcalc}}

To get the profile of the free membrane one has to solve the
appropriate Euler-Langrange (``shape'') equation. This equation is
typically a fourth order nonlinear partial differential equation
and thus in most cases impossible to solve analytically. One may,
however, consider cases where the membrane is indented only a
little and gradients are small. In that case one may linearize the
energy functional. In the constant indentation ensemble one gets
for the free part
\begin{equation}
  E = \int_{\Sigma_{\text{free}}} \!\!\!\! \romd A_{\|} \;
    \Big[ \frac{\kappa}{2}(\VECnab^{2}h) + \frac{\sigma}{2}(\VECnab h)^{2}\Big]
  \; ,
  \label{eq:energyfunctionalconstantheight}
\end{equation}
where $\romd A_{\|}$ is the area element on the flat reference
plane and $\Sigma_{\text{free}}$ is the projected surface of the
free pore-spanning membrane. The symbol $\VECnab$ denotes the
two-dimensional nabla operator in the reference plane.

The appropriate shape equation can be derived by setting the first
variation of energy~(\ref{eq:energyfunctionalconstantheight}) to
zero, yielding
\begin{equation}
  \VECnab^{2}\big(\VECnab^{2}-\lambda^{-2}\big)h = 0 \; .
  \label{eq:shapeequationlinear}
\end{equation}
The solution to this equation is a linear combination of the
eigenfunctions of the Laplacian corresponding to the eigenvalues 0
and $\lambda^{-2}$. For axial symmetry it is given by $h(\rho) =
h_{1} + h_{2}\ln{(\rho/\lambda)} + h_{3}I_{0}(\rho/\lambda) +
h_{4} K_{0}(\rho/\lambda)$, where $I_{0}$ and $K_{0}$ are the
modified Bessel functions of the first and the second kind,
respectively \cite{Abramowitz}. The constants $h_{1},
\ldots,h_{4}$ are determined from the appropriate boundary
conditions (see Appendix~\ref{app:boundaryconditions}):
\begin{subequations}
\label{eq:boundaryequationslinear}
\begin{eqnarray}
  h(\Rp) & = & 0 \; , \quad h(c) = -h_{0} + \frac{c^{2}}{2 \Rt} \; ,
  \label{eq:boundeqlin1}
  \\
  h'(\Rp) & = & 0 \; , \quad h'(c) = \frac{c}{\Rt} \; ,
  \label{eq:boundeqlin2}
  \\
  \text{and} \quad h''(c) & = & \frac{1}{\Rt} - \sqrt{\frac{2w}{\kappa}}
  \label{eq:boundeqlin3}
  \; ,
\end{eqnarray}
\end{subequations}
where a dash denotes a derivative with respect to $\rho$.  Even
though the differential equation is of fourth order, \emph{five}
conditions are required due to its moving boundary nature, \ie,
$c$ is to be determined from an adhesion balance---which is in
fact the origin of the fifth condition (\ref{eq:boundeqlin3}) 
(see Appendix~\ref{app:boundaryconditions}).

The solution of the boundary value
problem~(\ref{eq:shapeequationlinear},\ref{eq:boundaryequationslinear})
can be used in two ways to calculate the force for a prescribed
indentation: first, one can insert the profile back into the
functional (\ref{eq:energyfunctionalconstantheight}) to obtain the
energy of the equilibrium solution, which will then parametrically
depend on the indentation $h_{0}$. Its derivative with respect to
$h_{0}$ yields the force $F$. Second, one can also consider
stresses: in analogy to elasticity theory \cite{LaLi_elast} $F$ is
given by the integral of the flux of surface stress
\cite{surfacestresstensor} through a closed contour around the
tip.

The second approach is used in the present work; it has the
advantage that the final expression for the force can be written
in a closed form \cite{Steltenkamp_etal06} (see also
Appendix~\ref{app:diffgeostress}):
\begin{equation}
  F = 2 \pi \Rp \times \kappa \, \frac{\partial K}{\partial \rho}\Big|_{\rho=\Rp}
  \ .
  \label{eq:forcelinear}
\end{equation}
This equation is \emph{exact}.  Inserting the solution $h(\rho)$
of the boundary value
problem~(\ref{eq:shapeequationlinear},\ref{eq:boundaryequationslinear})
into (\ref{eq:forcelinear}) yields the value of the force in the
linear regime.

A little warning might be due here:
expression~(\ref{eq:forcelinear}) is evaluated at the rim of the
pore where the profile is flat even for high indentations.  One
might thus wonder whether inserting the small gradient solution
would actually lead to an exact result. This is, however, not the
case, because the membrane shape at the rim predicted by the
linear calculation is not identical with the prediction from the
full nonlinear theory---except for its flatness, which is enforced
by the boundary conditions. There is no magical way to avoid
solving the nonlinear shape equation if one wants the exact
answer.


\subsection{Complete nonlinear formulation\label{subsec:nonlinearcalc}}

Let us now shift to the angle-arclength parametrization and
consider the full nonlinear problem. In principle, the constant
height ensemble could be used here as well. It is, however,
technically much easier to fix $F$ instead in order to reduce the
number of boundary conditions one has to fulfill at the rim of the pore 
(see below and Appendix~\ref{app:numericalmethod}).

In this paragraph all variables with a tilde are scaled with $\pi
\kappa$, \ie: $\tilde{E}:=E/(\pi\kappa)$,
$\tilde{F}:=F/(\pi\kappa)$, etc. The energy functional of the free
membrane can then be written as
\cite{Berndl91,Hamiltonian95,Seifert97,wrapping2,Smith03}:
\begin{eqnarray}
  \tilde{E}
    & = & \int_{\underline{s}}^{\bar{s}} \romd s \; \tilde{L}
    = \int_{\underline{s}}^{\bar{s}} \romd s \;
    \Big\{ \rho \; \Big(\dot{\psi} +\frac{\sin{\psi}}{\rho} \Big)^{2}
      + \frac{2 \rho}{\lambda^{2}}
    \nonumber \\
    && \quad + \, \lambda_{\rho} (\dot{\rho} - \cos{\psi})
      + \lambda_{z} (\dot{z} - \sin{\psi})
      - \tilde{F} \dot{z} \Big\}
  \; ,\; \; \; \;
\end{eqnarray}
where $\underline{s}$ is the arclength at the contact point $c$ and $\bar{s}$
the arclength at $\Rp$. The dot denotes the derivative with respect to $s$.
The Langrange multiplier functions $\lambda_{\rho}$ and $\lambda_{z}$ 
ensure that the geometric conditions $\dot{\rho} = \cos{\psi}$ and 
$\dot{z} = \sin{\psi}$ are fulfilled everywhere.

In order to make the numerical integration easier let us rewrite
the problem in a Hamiltonian formulation
\cite{Berndl91,Hamiltonian95,wrapping2,Smith03}: the conjugate
momenta are $p_{\psi} = \partial \tilde{L}/\partial \dot{\psi} = 2
\rho [\dot{\psi} +\sin(\psi)/\rho]$, $p_{\rho} = \partial
\tilde{L}/\partial \dot{\rho} = \lambda_{\rho}$, and $p_{z} = \partial
\tilde{L}/\partial \dot{z} = \lambda_{z}-\tilde{F}$. The (scaled)
Hamiltonian is then given by
\begin{eqnarray}
  \tilde{H}\!& = &\!\dot{\psi} p_{\psi} + \dot{\rho} p_{\rho} 
    + \dot{z} p_{z} - \tilde{L}
  \nonumber \\
  & = &\!\frac{p_{\psi}^{2}}{4 \rho}
    - \frac{p_{\psi} \sin{\psi}}{\rho}
    - \frac{2 \rho}{\lambda^{2}}
    + p_{\rho} \cos{\psi} + (p_{z} + \tilde{F}) \sin{\psi}
  .\; \; \; \; \,
  \label{eq:Hamiltonian}
\end{eqnarray}
Note that $\tilde{H}$ is not explicitly dependent on $s$ and is thus a conserved
quantity. Instead of one fourth order one then has six first order ordinary
differential equations, the Hamilton equations:
\begin{subequations}
\label{eq:shapeequationsnonlinear}
\begin{alignat}{2}
  \dot{\psi} & \; = & \frac{\partial \tilde{H}}{\partial p_{\psi}}
    & \; = \; \frac{p_{\psi}}{2\rho} - \frac{\sin{\psi}}{\rho}
  \\
  \dot{\rho} & \; = & \frac{\partial \tilde{H}}{\partial p_{\rho}}
    & \; = \; \cos{\psi}
  \\
  \dot{z} & \; = & \frac{\partial \tilde{H}}{\partial p_{z}}
    & \; = \; \sin{\psi}
  \\
  \dot{p}_{\psi} & \; = & -\frac{\partial \tilde{H}}{\partial \psi}
    & \; = \; \Big[ \frac{p_{\psi}}{\rho} - (p_{z} + \tilde{F}) \Big] \cos{\psi}
      + p_{\rho} \sin{\psi}
  \label{eq:Hamiltonequation3}
  \\
  \dot{p}_{\rho} & \; = & -\frac{\partial \tilde{H}}{\partial \rho}
    & \; = \; \frac{p_{\psi}}{\rho}
      \Big(\frac{p_{\psi}}{4 \rho} - \frac{\sin{\psi}}{\rho} \Big)
      + \frac{2}{\lambda^{2}}
  \\
  \dot{p}_{z} & \; = & -\frac{\partial \tilde{H}}{\partial z}
    & \; = \; 0
  \label{eq:Hamiltonequation4a}
  \; .
\end{alignat}
\end{subequations}
According to the last equation, $p_{z}$ has to be constant along 
the profile. Its value can be found by considering the integral over 
the flux of surface stress which has to equal the applied force. This 
implies that $p_{z}$ vanishes everywhere (see 
Appendix~\ref{app:diffgeostress}).

Equations~(\ref{eq:shapeequationsnonlinear}) can be solved numerically 
subject to the boundary conditions (see also 
Appendices~\ref{app:boundaryconditions} and \ref{app:numericalmethod}):
\begin{subequations}
\label{eq:boundaryequationsnonlinear}
\begin{eqnarray}
  \psi(\bar{s}) & = & 0 \; , \quad \psi(\underline{s})=\alpha \; ,
  \label{eq:boundeqnonlin1}
  \\
  \dot{\psi}(\underline{s}) & = &
    \frac{(\cos{\alpha})^{3}}{\Rt}-\sqrt{\frac{2w}{\kappa}} \; ,
  \label{eq:boundeqnonlin2}
  \\
  \text{and} \quad\tilde{H} & = & 0
  \label{eq:boundeqnonlin3}
  \; ,
\end{eqnarray}
\end{subequations}
where contact radius $c$ and contact angle $\alpha$ are connected
via $c=\Rt\tan{\alpha}$. The solution to
(\ref{eq:shapeequationsnonlinear},
\ref{eq:boundaryequationsnonlinear}) gives the indentation $h_{0}$
for some prescribed force $\tilde{F}$.


\section{Results\label{sec:results}}

This chapter will summarize the characteristic features of the
solution to the boundary value
problems~(\ref{eq:shapeequationlinear},
\ref{eq:boundaryequationslinear}) and
(\ref{eq:shapeequationsnonlinear},
\ref{eq:boundaryequationsnonlinear}). In addition, the theory will
be shown to be in accord with available experimental results.

We will introduce some additional variable rescaling in order to
make generalizations of the results easier: lengths will be scaled
with $\Rt$. We also define
\begin{equation}
  \tilde{\sigma} := \frac{\sigma \Rt^{2}}{\kappa} \quad, \quad
  \tilde{w} := \frac{2 w \Rt^{2}}{\kappa} \quad \text{and} \quad
  \tilde{f} := \frac{F \Rt}{\pi \kappa}
  \; .
\end{equation}
In a typical experiment the curvature of the tip is of the order
of ten nanometer (5--40 nm) and pore radii may lie between 30 and
200 nm \cite{personalcommStelt}. The bending rigidity of a fluid
membrane may vary between one and a hundred $\kT$
\cite{bendingrigidityfluidmembrane}. One expects a maximum surface
tension of the order of a few mN/m, which is approximately the
rupture tension for a fluid phospholipid bilayer
\cite{surfacetension}. A maximum value of the adhesion can be
found by assuming that a few $\kT$ per lipid is stored if membrane
and tip are in contact. One arrives at $w_{\text{max}}\approx
10\,\text{mJ}/\text{m}^{2}$. For the continuum theory to be valid
Eqns.~(\ref{eq:boundeqlin3}, \ref{eq:boundeqnonlin2}) imply that
$\sqrt{2w_{\text{max}} / \kappa} \lesssim 1/d$, where $d\approx
5\,\text{nm}$ is the bilayer thickness. This estimate yields 
approx.\ the same maximum value for $w_{\text{max}}$ as before 
since $\kappa$ is at most 100 $\kT$. 

Thus, $\tilde{\sigma}$ and $\tilde{w}$ can in principal vary
between 0 and $10^{3}$. Realistically, if we set $\Rt =
10\;\text{nm}$ and consider a typical fluid phospholipid bilayer
with $\kappa \simeq 20\,\kT$, $\tilde{\sigma}$ and $\tilde{w}$ are
of the order of 1. Furthermore, we will focus on a pore radius of
$\tildeRp=3$ in the following.

In order to understand, how adhesion energy modifies the
force-distance behavior, let us first briefly revisit the case
where there is no adhesion between tip and membrane
($\tilde{w}=0$).


\subsection{No adhesion between tip and membrane\label{subsec:noad}}

\begin{figure}
\begin{center}
\includegraphics[scale=0.53]{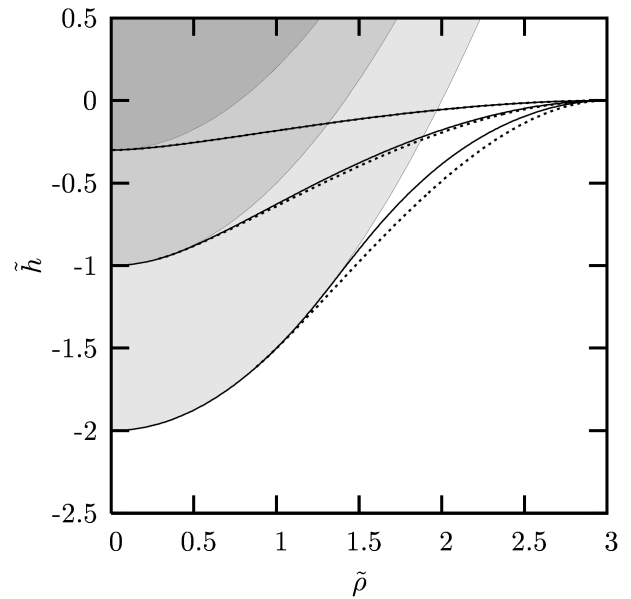}
\caption{Membrane profiles for different indentations
$\tilde{h}_{0}$, all for $\tilde{\sigma}=1$ (solid lines:
nonlinear calulations, dashed lines: linear approximation, grey
shades: AFM tips). The corresponding forces
$\tilde{f}(\tilde{h}_{0})$ for the three different indentations
are (nonlinear calculations): $\tilde{f}(0.3) = 0.81$,
$\tilde{f}(1) = 2.45$, $\tilde{f}(2) = 4.27$.}
\label{fig:profilesh0}
\end{center}
\end{figure}

In Fig.~\ref{fig:profilesh0} the shapes of the membrane for
different values of indentation are presented in scaled units. The
linear calculations are dotted whereas the exact result is plotted
with solid lines. For small indentations the two solutions
overlap; for increasing $\tilde{h}_{0}$, however, the deviations
become larger just as one expects for a small gradient
approximation (see also Ref.~\cite{wrapping2} for another
example).  While the differences are noticeable, they appear
fairly benign, such that one would maybe not expect big changes in
the force-distance behavior.  We will soon find out that these
hopes will not be fulfilled.

\begin{figure}
\begin{center}
\includegraphics[scale=0.24]{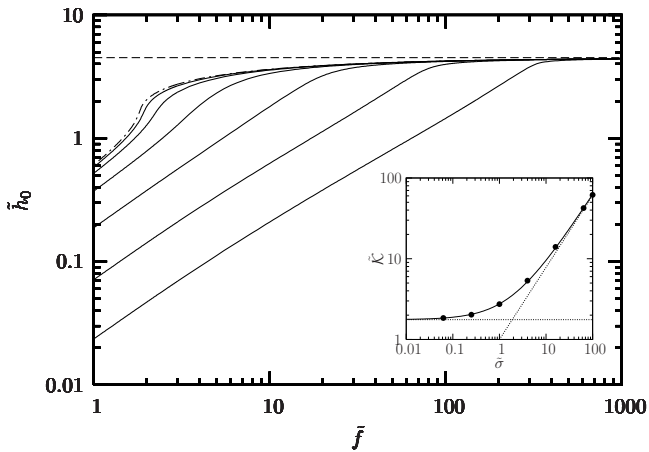}
\caption{Force-distance curves for $\tilde{w}=0$ and
$\tilde{\sigma}=\frac{1}{16},\frac{1}{4},1,4,16$ and $64$
($\sigma$ increasing from left to right). The curve for
$\tilde{\sigma}=0$ is dashed-dotted.  The inset shows the
corresponding scaled apparent spring constant $\tilde{\CALK}$ (see
Eqn.~(\ref{eq:CALK})) in the small force limit, illustrating its
two different regimes of small and large tension with a crossover
around $\tilde{\sigma}\simeq 1$.}\label{fig:1forcedistancew=0}
\end{center}
\end{figure}

\begin{figure}
\begin{center}
\includegraphics[scale=0.24]{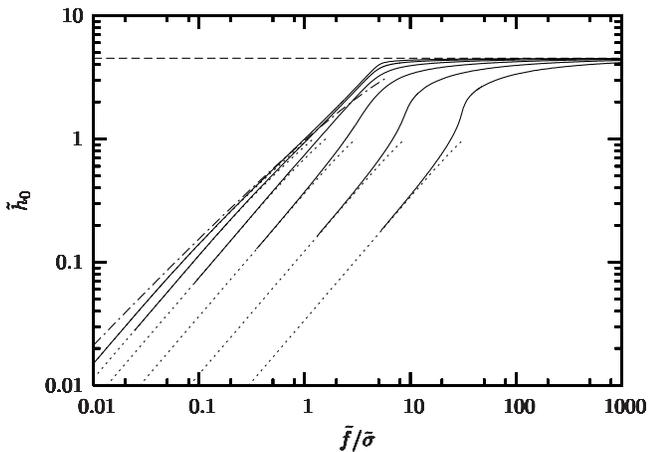}
\caption{Scaled force-distance curves for $\tilde{w}=0$ and
$\tilde{\sigma}=\frac{1}{16},\frac{1}{4},1,4,16$ and $64$
($\sigma$ increasing from right to left). The solution for
$\tilde{\sigma}\to\infty$ in the linear regime is dashed-dotted.
Nonlinear results are plotted with solid lines, the linear
approximation is dotted.}\label{fig:2forcedistancew=0}
\end{center}
\end{figure}

A deeper indentation also means that the tip has to exert a higher
force. In Figs.~\ref{fig:1forcedistancew=0} and
\ref{fig:2forcedistancew=0} log-log plots of force-distance curves
for different values of $\tilde{\sigma}$ are shown. The dashed
line marks the maximum indentation $\tilde{h}_{0,\text{max}} =
\tildeRp^{2} / 2$ which is allowed by the geometry of tip and
pore. In the limit of high forces all curves converge and approach
$\tilde{h}_{0,\text{max}}$; for small forces the curves are linear
in $\tilde{f}$.  Let us quantify the indentation response by
defining the (scaled) apparent spring constant $\tilde{\CALK}$ of
the nanodrum-AFM system via
\begin{equation}
\tilde{\CALK} =
\left.\frac{\partial\tilde{f}}{\partial\tilde{h_0}}\right|_{\tilde{w},\tilde{\sigma}}
\ . \label{eq:CALK}
\end{equation}
A linear force-distance-curve has a constant $\tilde{\CALK}$ and
thus follows an apparent Hookean behavior $\tilde{f} =
\tilde{\CALK}\tilde{h}_0$.  In unscaled units, the spring constant
is given by $\CALK=\partial F/\partial h_0 =
\tilde{\CALK}\pi\kappa/\Rt^2$.  For typical values $\Rt=10\,\text{nm}$
and $\kappa=20\,\kT$ this implies $\CALK=\tilde{\CALK}\times
2.6\,\text{mN/m}$.

The smaller $\tilde{\sigma}$, the less force has to be applied to
reach the same indentation (see Fig.~\ref{fig:1forcedistancew=0}).
For decreasing $\tilde{\sigma}$ the force-distance curves converge
to the limiting curve of the pure bending case, for which
$\tilde{\sigma}=0$; this is plotted dashed-dotted in
Fig.~\ref{fig:1forcedistancew=0}. In the opposite pure tension
limit ($\kappa\rightarrow 0$ or $\tilde{\sigma}\to\infty$) the
curves become essentially linear in $\tilde{\sigma}$, as can be
seen clearly after scaling out the tension (see
Fig.~\ref{fig:2forcedistancew=0}). It is possible to calculate
this second limiting curve in the linear regime: the linearized
Euler Lagrange equation reduces to the Laplace equation, $\Delta h
= 0$, which is solved by $h(\rho)=d_{1} + d_{2}\ln(\rho/\Rp)$ in
the present axial symmetry. The constants $d_{1}$ and $d_{2}$ can
be determined with the help of the two boundary conditions
$h(\Rp)=0$ and $h(c)=-h_{0}+c^{2}/2\Rt$. The contact point $c$ is
then determined by a straightforward energy minimization. The
final result for the indentation depth is:
\begin{equation}
  \tilde{h}_{0}^{\tilde{\sigma}\to\infty}
  = \frac{\tilde{f} / \tilde{\sigma}}{4}
    \bigg[1-\ln{\bigg(\frac{\tilde{f} / \tilde{\sigma}}{2\tildeRp^{2}}\bigg)}\bigg]
  \; ,
\end{equation}
which is plotted dashed-dotted in
Fig.~\ref{fig:2forcedistancew=0}. At any given penetration the
force is now strictly proportional to the tension. Notice also the
remarkably weak (logarithmic) dependency of penetration on pore
size.

\begin{figure}
\begin{center}
\includegraphics[scale=0.8]{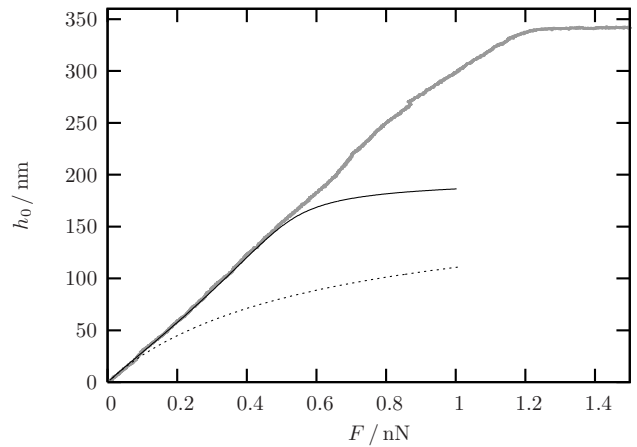}
\caption{Comparison between experiment (solid grey line) and
theory (dashed line: linear approximation; solid black line:
nonlinear calculations) \cite{Steltenkamp_etal06}. The theoretical
curves are obtained with the following parameters: $\Rp =
90\,\text{nm}$, $\Rt=20\,\text{nm}$, $\kappa = 10^{-19} \,
\text{J}\approx 24\,\kT$, $\sigma = 1.1 \,
\text{mN/m}$.}\label{fig:experiment}
\end{center}
\end{figure}

All force-distance curves presented in this section exhibit a
linear behavior for small forces.  In this limit the scaled spring
constant for the systems just discussed is well described by the
empirical relation $\tilde{\CALK}\simeq 1.76 +
\tilde{\sigma}^{0.89}$ (see inset in
Fig.~\ref{fig:1forcedistancew=0}). Combining this with our
observation that for typical system parameters
$\CALK=\tilde{\CALK}\times 2.6\,\text{mN/m}$, we see that a
nanodrum's stiffness can be very well matched by available (soft)
AFM cantilevers, showing that the suggested experiments are indeed
feasible.  In fact, Fig.~\ref{fig:experiment} shows the results of
such an indentation experiment (solid grey line). Here, a fluid
DOTAP (1,2-dioleoyl-3-trimethylammonium-propane chloride) membrane
was suspended over a pore of radius $\Rp=90\,\text{nm}$ and
subsequently probed with a tip of radius $\Rt=20\,\text{nm}$
\cite{Steltenkamp_etal06}.  The apparent spring constant is found
to be $3.9\,\text{mN/m}$. To fit the data we optimized the
material parameters $\sigma$ and $\kappa$. The linear
approximation (asymptotically) matches the curve down to an
indentation depth of about 40 nm as one can see in
Fig.~\ref{fig:experiment} (dashed line). For larger indentations
the small gradient assumption breaks down. The nonlinear
calculation (solid black line) describes the data correctly down
to a much deeper penetration depth of 150 nm but diverges for
larger values. This deviation is most likely \emph{not} a failure
of the elastic model but a consequence of our simplified
assumptions for the tip geometry. As shown in Fig.~1B of the
Supplementary Information to Ref.~\cite{Steltenkamp_etal06} the
tip is parabolic at its apex, but further up it narrows quicker
and assumes a more cylindrical shape. It therefore can penetrate
the pore much deeper than one would expect if the parabolic shape
were correct for the entire tip.

Apart from this difficulty, theory and experiment are in good
agreement. There is, however, a catch.  Since we cannot trust the
force-distance behavior close to the depth-saturation (due to its
displeasingly strong dependence on the actual tip shape), the
remaining interpretable part of the force-distance-curve is
linear, and its slope is the only parameter that can be extracted
from the data \cite{axisintercept}. For the theoretical
calculation one needs two parameters, $\sigma$ and $\kappa$.
Fitting both to a line is not possible. In
Ref.~\cite{Steltenkamp_etal06} this obstacle was overcome by
estimating $\kappa$ from other measurements to be about
$10^{-19}\,\text{J}$. The surface tension $\sigma$ could then be
adjusted to $1.1\,\text{mN/m}$ to match the data---which,
reassuringly, is a very meaningful value.

Alternatively, one may proceed in a different manner. In the
experiment a small snap-off peak could be observed upon retraction
of the AFM tip which was due to the attraction between tip and
membrane. Although this could be neglected in the interpretation
of the measurements of Ref.~\cite{Steltenkamp_etal06}, one may
think of deliberately increasing the adhesion between membrane and
tip in a follow-up experiment by chemically functionalizing the
tip. With this additional tuning parameter one may get further
information on the values of the material parameters in question.


\subsection{Including adhesion between tip and membrane\label{subsec:ad}}

In the following, we will also allow for adhesion between tip and
membrane, \ie\ $\tilde{w}$ is not necessarily equal to zero. This
will change the qualitative behavior of the force-distance curves
dramatically: for fixed $\tilde{\sigma}$ and $\tilde{w}$ different
solution branches can be found. A hysteresis may occur as well, as
we will see in the next section. Additionally, stable membrane
profiles exist even if the tip is pulled upwards. It is therefore
possible to calculate the maximum pulling force that can be
applied before the tip detaches from the membrane and relate it to
the value of the adhesion between tip and membrane.

\subsubsection{Weak adhesion energy\label{subsubsec:weakad}}

In this section, we will first investigate the case of weak
adhesion, $\tilde{w} \le 5$. The scaled surface tension
$\tilde{\sigma}$ will be fixed to 1.  It turns out that once the
tip is adhesive, ``overhang'' profiles may occur, \ie, shapes
where at some point $|\psi(s)|>90^\circ$.  We will first ignore
these solution branches and come back to them later.

\begin{figure}
\begin{center}
\includegraphics[scale=0.58]{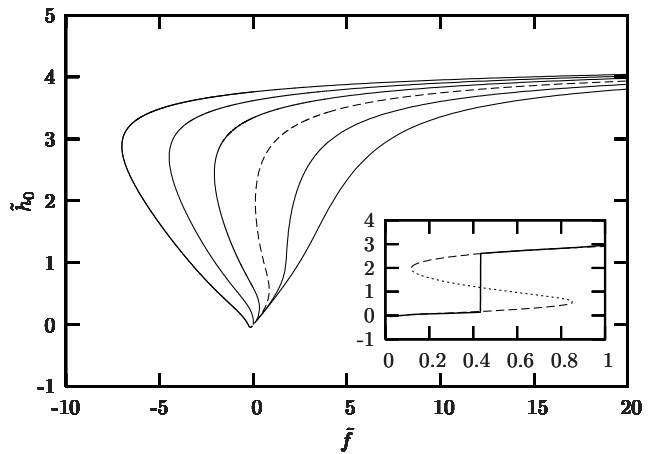}
\caption{Force-distance curves for $\tilde{\sigma}=1$ and
$\tilde{w}=0,1,2,3,4,5$ (from right to left). The region of
hysteresis in the curve for $\tilde{w}=2$ is magnified in the
inset. In this case the energy barrier at $\tilde{f}=0.414$ is
approx.\ $1\,\kappa$.  Overhang branches are omitted.}
\label{fig:forcedistancew1}
\end{center}
\end{figure}

Fig.~\ref{fig:forcedistancew1} illustrates force-distance-curves
for $\tilde{w}=0,1,\ldots,5$.  Compared to the nonadhesive case,
for which an essentially linear behavior levels off towards
maximum penetration, adhesive tips behave quite differently.
Already for $\tilde{w}=1$ an initial Hookean response at small
forces is soon followed by a regime in which the system displays a
much greater sensitivity towards an externally applied stress,
\ie, where the scaled spring constant $\tilde{\CALK}$ drops at
intermediate penetrations. Physically this of course originates
from the fact that adhesion \emph{helps} to achieve higher
penetrations, because the tip is pulled towards the membrane, but
notice that this does not lead to a uniform reduction of
$\tilde{\CALK}$: softening only sets in beyond a certain
indentation.

Shortly beyond $\tilde{w}=1$ a point is reached where the
force-distance-curve displays a vertical slope at which the
apparent spring constant $\tilde{\CALK}$ vanishes.  For even
larger values of adhesion a hysteresis loop opens, featuring a
locally unstable region with $\tilde{\CALK}<0$.  This is the case
for $\tilde{w}=2$, and the region around the instability is
magnified in the inset of Fig.~\ref{fig:forcedistancew1}. Notice
that the dotted branch corresponding to $\tilde{\CALK}<0$ still
belongs to solutions for which the functional (\ref{eq:Etotal}) is
stationary, yet the energy plotted against penetration
$\tilde{h}_0$ (or, alternatively, contact angle $\alpha$) has a
local \emph{maximum}, confirming that these solutions are unstable
against contact point variations.  The two dashed branches in the
inset of Fig.~\ref{fig:forcedistancew1} have a positive
$\tilde{\CALK}$ and correspond to local minima in the energy,
however, they are \emph{globally} unstable against the alternative
minimum of larger or smaller $\tilde{h}_0$.  The true global
minimum is indicated by the bold solid curve, which exhibits a
discontinuity at $\tilde{f}=0.414$.

Depending on the current scanning direction this hysteretic
force-distance-curve manifests itself in a snap-on or snap-off
event. Such a behavior is reminiscent of a \emph{buckling}
transition (such as for instance Euler buckling of a rod under
compression \cite{LaLi_elast})---with two caveats: first, notice
that the membrane does \emph{not} stay flat up to a critical
buckling force at which it suddenly yields; rather, the system
starts off with a linear stress-strain relation and only later
undergoes an adhesion-driven discontinuity.  Appreciating this
point is quite important for the interpretation of measured
force-distance curves:  upon approach of tip and membrane the
snap-on will occur \emph{neither} at zero force \emph{nor} at zero
penetration. Second, one should not forget that hysteresis is
ultimately a consequence of the energy barrier which goes along
with such discontinuities. For macroscopic systems this barrier is
typically so big that the transition actually happens at either of
the two end-points of the S-shaped hysteresis curve, where the
barrier vanishes (the ``spinodal points'').  However, for
nano-systems barriers are much smaller, comparable to thermal
energy $\kT$, such that thermal fluctuations can assist the
barrier-crossing event.  In the present case the barrier at the
equilibrium transition point is about $1\,\kappa$, \ie, about
$20\,\kT$ for typical bilayers. However, already at
$\tilde{f}=0.5$ its magnitude has decreased by about 20\%. 
This shows that we have to expect a narrowing-down of the 
hysteresis amplitude compared to an athermal buckling scenario.

Upon increasing the adhesion $\tilde{w}$ even further, one will
reach a critical value $\tilde{w}_0$ at which the ``back-bending
branch'' of the force-distance-curve touches the vertical line
$\tilde{f}=0$.  At this point the tip is being pulled into the
pore even if there is no force at all.  Conversely, neglecting
barrier complications, this also implies that at the critical
adhesion energy $\tilde{w}=\tilde{w}_{0}$ an \emph{infinitesimal}
pulling force will suffice to unbind tip and membrane, \emph{even
though} the adhesion between tip and membrane is greater than
zero. It is very important to keep this fact in mind if one wants
to use AFM measurements for the determination of adhesion
energies.

For $\tilde{w}>\tilde{w}_0$ one obtains stable solutions even when
pulling the tip upwards (where $\tilde{f}<0$) \cite{ftildele0}.
The maximum possible force before detachment,
$\tilde{f}_\text{det}$, again corresponds to the leftmost point of
the back-bend, and it increases with increasing $\tilde{w}$; we
will come back to this later.  Notice that detachment always
happens for values of $\tilde{h}_0$ which are \emph{positive},
\ie, when the AFM tip is still \emph{below} the substrate level.
Contrary to what one might have expected, pulling will in this
case \emph{not} draw the membrane upwards into a tubular lipid
bilayer structure (a ``tether''), which at some specific
elongation will fall off from the tip and snap back.  Rather, the
strong adhesion pulls the tip far into the pore, and while pulling
on it indeed lifts it up, unbinding still happens below pore rim
level.


\subsubsection{Strong adhesion energy\label{subsubsec:strongad}}

\begin{figure*}
\begin{center}
\includegraphics[scale=1.0]{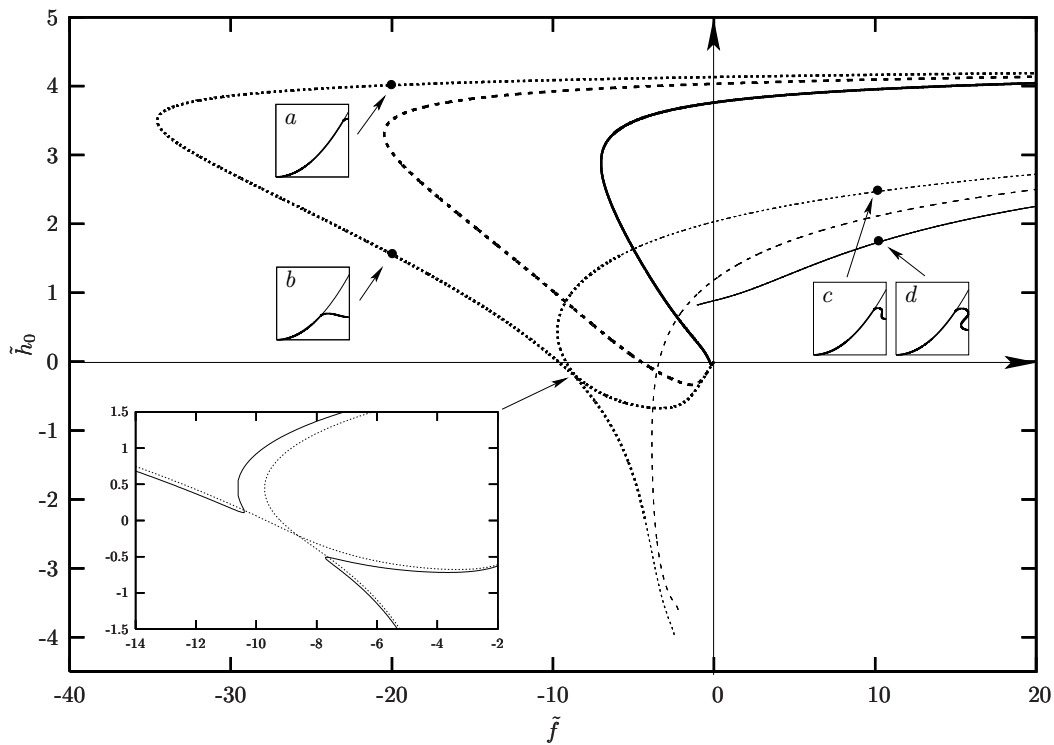}
\caption{Force-distance curves for $\tilde{\sigma}=1$ and
$\tilde{w}=5$ (solid), 10 (dashed) and 15 (dotted). A thinner line
style is used for those parts of the curves where the
corresponding profiles exhibit overhangs. In the insets
\emph{a}--\emph{d} profiles for different values of
$(\tilde{f},\tilde{h}_{0})$ are depicted (scaling is
$\tilde{h}:\tilde{\rho}=5:3$).  In the inset on the lower left
corner the ``branch splitting'' is shown as discussed in the text
($\tilde{w}=15.0$ (dotted line) and $ 15.5$ (solid
line)).}\label{fig:forcedistancew2}
\end{center}
\end{figure*}

At even larger adhesion energy entirely new stationary solution
branches emerge, as Fig.~\ref{fig:forcedistancew2} illustrates for
$\tilde{w}\in\{5,10,15\}$ and $\tilde{\sigma}=1$. We first
recognize the well-known hysteretic branch, already seen in
Fig.~\ref{fig:forcedistancew1}, which for increasing $\tilde{w}$
extends to much larger negative forces, even though the snap-off
height $\tilde{h}_0$ does only change marginally.  The shapes of
two typical profiles are illustrated in the insets \emph{a} and
\emph{b}.  Notice that this branch is always connected to the
origin, but for larger values of $\tilde{w}$ it starts off into
the third quadrant (negative values for $\tilde{f}$ and
$\tilde{h}_0$). At first sight it seems that we finally get
solutions which correspond to a pulled-up membrane; however, this
region close to the origin corresponds to a maximum and is thus
unstable.

\paragraph{Overhang branches.}

Contrary to the hysteretic branches, the new branches depicted in
Fig.~\ref{fig:forcedistancew2} do \emph{not} connect to the
origin.  This classifies them as a genuinely nonlinear phenomenon,
since they cannot be obtained as a small perturbation around the
state $\tilde{f}=\tilde{h}_0=0$. In the first quadrant
($\tilde{f},\tilde{h}_0>0$) they all correspond to profiles which
show overhangs (see inset $c$ and $d$).  These branches had been
omitted in Fig.~\ref{fig:forcedistancew1}, since for weak adhesion
they always correspond to maxima and are thus irrelevant. This
changes for stronger adhesion, though, where they become stable in
certain regions (for instance, inset $c$ is locally stable). The
details by which this happens are complicated and will be
discussed in more detail below.

Following the new branches to negative forces we see that the one
for $\tilde{w}=15$ loses its overhang around $\tilde{f}\approx-6$.
That this can happen continuously is not surprising, since within
angle-arc-length parametrization there is nothing special about
the point where $|\psi|=90^\circ$ (only the shooting method might
use occurrences of $|\psi|>90^\circ$ as a potential termination
criterion for integration).

\paragraph{Branch splitting.}

We also see that (for sufficiently large $\tilde{w}$) there is a
point where the hysteresis branch \emph{intersects} the new
nonlinear branch.  There the values of $\tilde{f}$ and
$\tilde{h}_0$ coincide for both branches, but the detachment angle
$\alpha$ and the total energy of the profile are generally
different.  However, the difference in energy at the intersection
decreases with increasing $\tilde{w}$, and around $\tilde{w}=15.3$
it finally vanishes.  At this degenerate point a \emph{branch
splitting} occurs, where the connectivity of the two branches
re-bridges, as illustrated in the lower left inset in
Fig.~\ref{fig:forcedistancew2}.  Rather than connecting to the
origin, the wide loop of the original hysteresis branch now joins
into the overhang branch of the first quadrant, while its bit that
was connected to the origin now joins into the overhang branch in
the third quadrant.

\paragraph{Cusps.}

\begin{figure*}
\begin{center}
\includegraphics[scale=1.0]{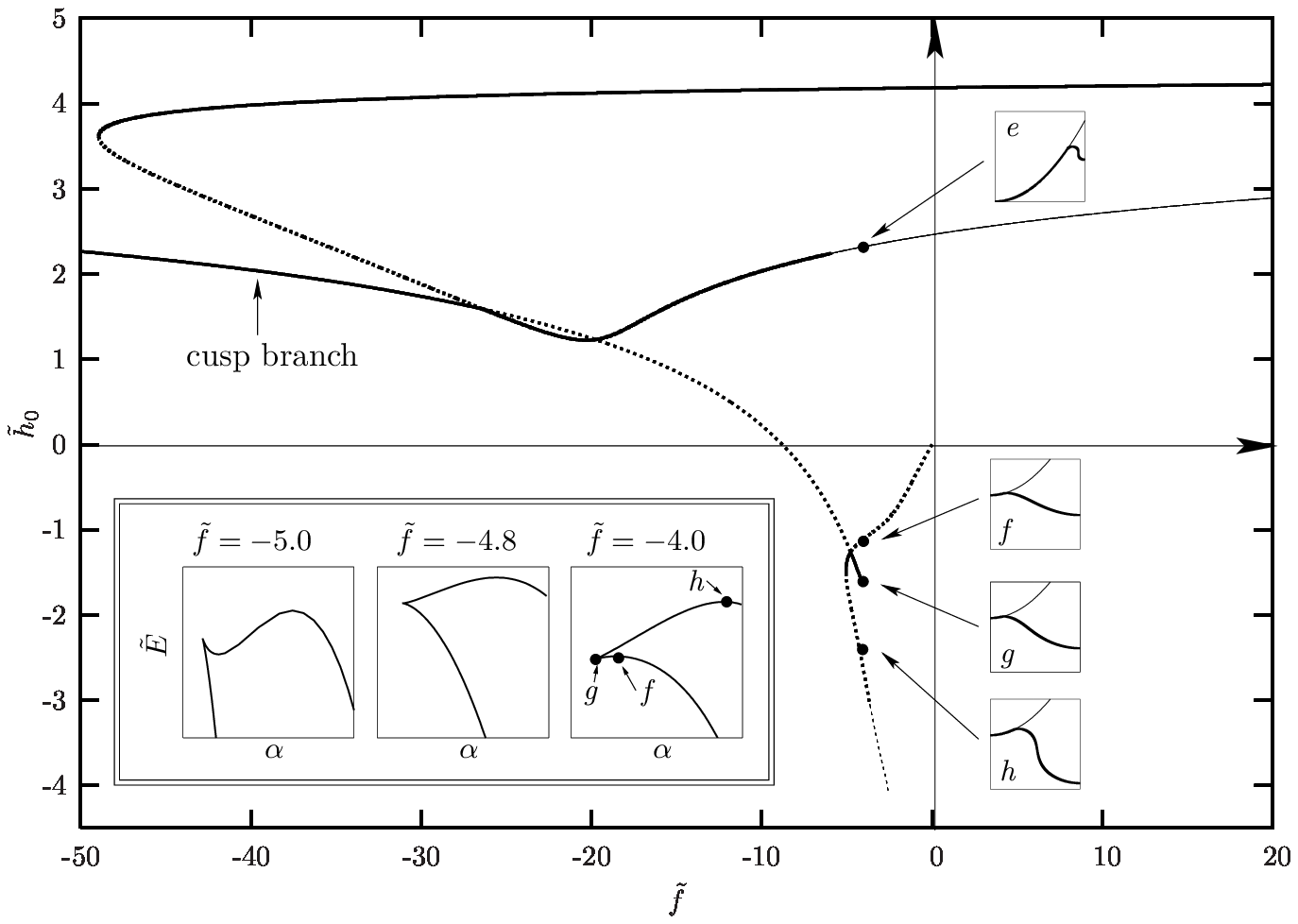}
  \caption{Force-distance curves for $\tilde{\sigma}=1$ and $\tilde{w}=20$ 
  (including cusp branch). Solid lines correspond to local minima, dotted 
  lines to local maxima. A thinner line style is used for those parts of the 
  curves where the corresponding profiles exhibit overhangs. In the insets \emph{e} - 
  \emph{h} profiles for different values of $(\tilde{f},\tilde{h}_{0})$ are 
  depicted (scaling is $\tilde{h}:\tilde{\rho}=5:3$).
  In the inset on the left lower corner the total energy $\tilde{E}$ is 
  plotted as a function of detachment angle $\alpha$ for different forces 
  $\tilde{f}$ 
  (see text for further explanation).}
  \label{fig:forcedistancew3}
\end{center}
\end{figure*}

Figure~\ref{fig:forcedistancew3} shows the force-distance curve
branches for the even larger adhesion energy $\tilde{w}=20$. This
depicts a situation well after the branch-splitting, so we
recognize the old hysteretic branch connecting with overhangs, and
the branch connecting to the origin extending exclusively in the
third quadrant.  In contrast to Fig.~\ref{fig:forcedistancew2},
the line styles in Fig.~\ref{fig:forcedistancew3} are chosen to
illustrate local minima (solid) or maxima (dotted).  What
immediately strikes one as surprising is that the profiles at
$\tilde{f}=-4$ belonging to the insets $f$ and $h$ \emph{both}
correspond to maxima, even though they sit on both sides of a
back-bending branch, close to its end (compare this to the
``usual'' scenario at ($\tilde{f}\approx -49$, $\tilde{h}_0\approx
3.6$). Moreover, the solution belonging to inset $f$ turns into a
local minimum for slightly more negative forces, \emph{without}
any noticeable features of the branch.  How can this happen?

The explanation is illustrated in the lower left inset in
Fig.~\ref{fig:forcedistancew3}, which shows the total energy as a
function of detachment angle $\alpha$.  Recall that extrema in
this plot correspond to stationary solutions.  As can be seen, the
energy is \emph{multivalued}, meaning that there exists more than
one solution at a given detachment angle (these would then also
differ in their value of their penetration $\tilde{h}_0$). But
more excitingly, this graph exhibits a \emph{boundary extremum} at
a lowest possible nonzero value of $\alpha$ in the form of a
\emph{cusp}.  This is how one can have two successive maxima on a
curve without an intervening minimum---the minimum is simply not
differentiable.  Hence, there is a \emph{third} solution branch,
corresponding to the cusp, at which the contact curvature
condition from Eqn.~(\ref{eq:boundeqnonlin2}) is \emph{not}
satisfied, because this condition is blind to the possibility of
having non-differentiable extrema.  Plotting this cusp branch also
into Fig.~\ref{fig:forcedistancew3}, we finally understand how the
switching of a maximum into a minimum happens:  it occurs at the
point of intersection with the cusp branch.  As the lower left
inset in Fig.~\ref{fig:forcedistancew3} illustrates, the maximum
belonging to the solution $f$ joins the cusp-minimum (belonging to
solution $g$) in a \emph{boundary flat point}, roughly at force
$\tilde{f}=-4.8$.  For more negative forces this flat point turns
up, leaving a boundary cusp \emph{maximum} and a new
differentiable minimum.  Notice that a similar exchange happens
once more at ($\tilde{f}\approx -26.4$, $\tilde{h}_0\approx 1.6$).
Incidentally, since at the cusp the contact curvature condition is
not satisfied, and since this is the only point where the adhesion
energy $\tilde{w}$ enters, the location and form of the cusp
branch is \emph{independent} of the value of $\tilde{w}$.

The existence of the cusp branch poses the question, whether the
solutions corresponding to it are physically relevant (at least
the ones which are minima). It is not so much the lack of
differentiability at a cusp minimum which causes concern, but
rather the fact that it is located at a \emph{boundary}.  Take for
instance the $\tilde{E}(\alpha)$ curve in the lower left inset of
Fig.~\ref{fig:forcedistancew3} corresponding to $\tilde{f}=-4$.
Now consider a (nonequilibrium) solution which sits on the upper
branch, somewhere between the solutions $g$ and $h$.  To lower the
energy, this solution will reduce the detachment angle $\alpha$,
thereby approaching the minimum at $g$.  But once $g$ has been
reached, no further reduction in $\alpha$ seems possible, since
for smaller values no equilibrium solution exists.  The crucial
point is that our present theory is insufficient to answer what
\emph{else} would be going on for smaller $\alpha$. It could for
instance be that there are indeed solutions, \emph{but they are
not time-independent}.  This might be analogous to the well-known
situation of a soap film spanned in the form of a catenoidal
minimal surface between two coaxial circular rings of equal radius
$R$. It is easy to show that for a ring separation exceeding
$1.325\,R$ no more stationary solution exists, even though the
limiting profile is in no way singular \cite{Arfken}. However,
when slowly pulling the two rings beyond this critical separation,
the soap film does not suddenly rupture. Rather, it becomes
\emph{dynamically} unstable and begins to collapse.  In the case
we are studying here, the system drives itself to the singular
boundary point, and without a truly dynamical treatment it is not
possible to conclude whether it would remain there or start to
dynamically approach a different solution.  For this reason we do
not want to overrate the significance of the cusp branch; yet, its
existence is still important in order to explain the behavior of
the other ``regular'' branches, for instance their metamorphosis
from maximum-branches into minimum-branches or vice versa.

\paragraph{Detachment forces.}

\begin{figure}
\begin{center}
\includegraphics[scale=0.80]{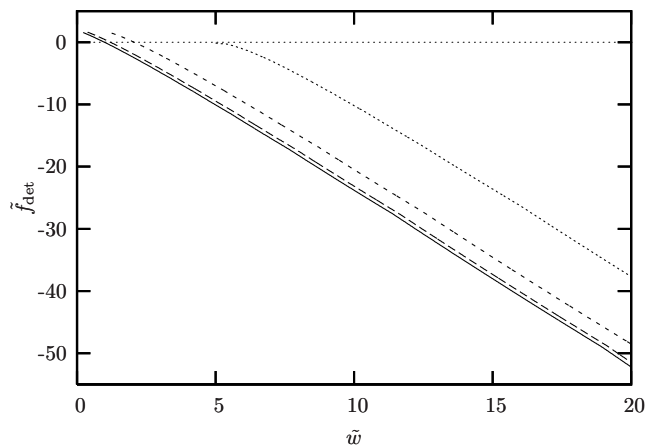}
\caption{Scaled detachment force $\tilde{f}_{\text{det}}$ as a
function of scaled adhesion energy $\tilde{w}$ for four values of
the scaled tension, $\tilde{\sigma} = \frac{1}{16}$,
$\frac{1}{4}$, $1$ and $4$ ($\tilde{\sigma}$ increasing from left
to right).}\label{fig:detachmentforce}
\end{center}
\end{figure}

A measurable quantity in the experiment is the detachment force
between tip and membrane, which is the maximum applicable pulling
force $\tilde{f}_{\text{det}}$ before the tip detaches from the
membrane. In Fig.~\ref{fig:detachmentforce} this force is plotted
as a function of adhesion energy $\tilde{w}$ for different values
of the scaled tension $\tilde{\sigma}$. Starting from a certain
threshold adhesion $\wthr(\tilde{\sigma})$, below which no
hysteresis occurs, $\tilde{f}_{\text{det}}$ decreases with
increasing $\tilde{w}$ and exhibits a linear behavior for higher
adhesions. Increasing $\tilde{\sigma}$ also increases the
threshold adhesion (\eg\ $\wthr(1/16)=0.22$ compared to
$\wthr(1)=1.25$).  In the large-$\tilde{w}$-limit
$\tilde{f}_{\text{det}}/\tilde{w}=F_{\text{det}}/2\pi Rw$ finally
approaches a limit which is independent of $\kappa$ and $\sigma$
and only depends on the geometry.  The elasticity of the membrane
no longer influences the measurement of the adhesion energy -- not
because the membrane is not deformed, but rather because its
deformation energy is subdominant to adhesion.  But for more
realistic smaller values of $\tilde{w}$ this decoupling does not
happen, and adhesion energies can only be inferred from the
detachment force when a full profile calculation is performed.

At higher values of $\tilde{\sigma}$ also other qualitative
features (such as additional instabilities) occur. However, these
ramifications will not be discussed in the present work.

\paragraph{Tethers.}

Characteristically, the detachment happens at deep indentations
($\tilde{h}_{0}$ close to the maximum indentation possible).  Long
pulled-out membrane tubes (``tethers''), as they have been studied
in the literature \cite{Goldstein02,Derenyi02,Koster05}, are not
observed.  Even though in our calculations we find profiles with
$\tilde{h}_{0}<0$, these solutions either correspond to energetic
maxima, or they are only \emph{local} minima -- with the global
minimum at $\tilde{h}_0>0$ corresponding to a significantly lower
energy. This is a consequence of the adhesion balance present in
our situation: upon pulling upwards, it is more favorable for the
tip either to be ``sucked in'' completely or to detach from the
membrane, rather than forming a long tether. As
Fig.~\ref{fig:forcedistancew3} shows, there is a very small
``window of opportunity'' at $\tilde{f}\approx -5$ where (locally)
stable solutions pulled above the surface exist.  Yet, their
profiles look essentially like the ones of inset $f$ or $g$ and
show no resemblance to real long tethers.  Upon increasing the
force they become unstable, such that the tip either falls of the
membrane, or is drawn below the membrane plane (notice that there
exist two minima at $\tilde{f}$ slightly smaller than $-5$, but
both at positive indentation).

This analysis shows that it appears impossible to pull tethers
using a probe with a certain binding
energy, despite existing experiments in which tethers of
micrometer size were generated \cite{DaiShe95_99,MaeSenMeg02,Koster05}. 
Consequently, the assumption of an adhesion balance does not seem 
to be correct in these cases. 
Indeed, in these studies the experimental setup was different 
(membrane-covered micron-sized beads \cite{DaiShe95_99,Koster05} and 
AFM tips covered by lipid multilayers \cite{MaeSenMeg02}). 
In the present situation tethers are also observed 
\cite{Steltenkamp_etal06}, but these events are not very reproducible, 
and based on the above calculations we would tentatively attribute 
them to a pinning of the membrane at some irregularity of the tip.


\section{Discussion\label{sec:discussion}}

In the previous sections we have discussed the indentation of a pore-spanning 
bilayer by an AFM tip. We have seen that the force-distance curves show a linear 
behavior for small forces in a broad parameter range if the adhesion between 
tip and membrane vanishes. Even though this is in agreement with recent 
experiments (\cite{Steltenkamp_etal06}, see also Fig.~\ref{fig:experiment}), 
such a linear behavior is unfortunately too featureless to reveal the values 
of both elastic material constants, $\sigma$ and $\kappa$. 

One way out of this apparent cul-de-sac would be to repeat the experiment for 
different pore radii $\Rp$ while keeping all other parameters fixed. Since 
$\sigma$ and $\kappa$ are the same for all pore sizes in that case, it should be 
possible to extract their value from the measured force-distance curves. 
Note that one does not have to fit both parameters simultaneously if one at first 
considers a pore where the radius is much larger than the characteristic 
lengthscale $\lambda$ (see Eqn.~(\ref{eq:characteristiclengthscale})). The 
corresponding system is in the high tension regime and the force-distance 
relation only depends on the surface tension (see Section~\ref{subsec:noad}). 
After determining $\sigma$ from the resulting curve, one may subsequently extract 
the value of $\kappa$ from a measurement of a system with smaller pore size. 

The elastic constants can also be obtained by considering systems where the 
adhesion $w$ between tip and membrane has been increased experimentally. As we 
have seen, the curves change their behavior dramatically for $w\ne 0$. It should 
thus be possible to fit two parameters to the resulting curves which would yield 
a local $\kappa$ and $\sigma$ in one fell swoop whereas $w$ can simultaneously 
be determined from the snap-on of the tip upon approach to the bilayer.
The experimentalist, however, has to make sure in that case that the line of
contact between tip and membrane is really due to a force balance as described
in this paper and not due to other effects such as pinning of the membrane to
single spots on the tip. In practice, this is rather difficult and will be a
challenge for future experiments.

One also has to keep in mind that the assumption of a perfect parabolic tip is 
quite simplistic compared to the experimental situation. It is probably valid 
in the vicinity of the apex but generally fails further up. Since the 
force-distance behavior close to the depth-saturation depends strongly on the 
actual tip shape, one can only use that part of the force-distance curve for 
data interpretation where the indentation is still small.
To predict the whole behavior the exact indenter shape has to be known: as long 
as the situation stays axisymmetric one may, in principle, redo the calcutions 
of this publication with the new shape. This is, however, rather tedious and 
therefore inexpedient in practice. 

Our theorical approach does not account for hydrodynamic effects although the
whole setup is in water and the AFM tip is moved with a certain velocity. First 
measurements have shown, however, that it is possible to increase the velocity
of the tip up to 60\,\muem$\text{s}^{-1}$ without altering the force-distance 
curves dramatically \cite{Steltenkamp_etal06}. One can understand this result 
with the help of the following simple estimate: assume that the tip is a sphere 
of radius $\Rt$ moving with the velocity $v$ in water. When indenting the 
membrane to a distance $d$ it will also have to overcome a dissipative 
hydrodynamic force $F_{\text{diss}}$ in addition to the elastic resistance of 
the membrane. The energy dissipated in this process, 
$E_{\text{diss}}=F_{\text{diss}}d=6\pi\eta\Rt v d$, is of the order of the 
thermal energy if typical values are inserted 
($\eta=10^{-3} \text{Pa s}$, $\Rt=10\,\text{nm}$, $v=$ 60\,\muem$\text{s}^{-1}$, 
$d=100\,\text{nm}$). This is substantially smaller than the corresponding 
elastic energy $E_\text{elast}$. Complications arising from a correct 
hydrodynamical treatment were thus omitted here. 

Including adhesion, the velocity of the measurement should nevertheless be as
slow as possible to ensure that the line of contact equilibrates due to the
force balance. If this is guaranteed, one can also check whether the predicted
linear behavior between detachment force and adhesion is actually valid.

\begin{acknowledgments}

We thank Siegfried Steltenkamp and Andreas Janshoff for providing the
experimental data (see Fig.~\ref{fig:experiment}). We have greatly benefitted
from discussions with them and with Jemal Guven.
MD acknowledges financial support by the German Science Foundation through grant
De775/1-3.

\end{acknowledgments}


\appendix

\section{Boundary conditions \label{app:boundaryconditions}}

In this appendix we will explain the origin of the
boundary conditions~(\ref{eq:boundaryequationslinear}) and
(\ref{eq:boundaryequationsnonlinear}):
Eqns.~(\ref{eq:boundeqlin1}) follow simply from the requirement of continuity at
the pore rim and the point where the membrane leaves the tip. Asking for a
membrane that has no kinks and thus no diverging bending energy gives
Eqns.~(\ref{eq:boundeqlin2}) and (\ref{eq:boundeqnonlin1}).

If the membrane is free to choose its point of detachment as it is assumed here,
an adhesion balance at the tip yields another boundary condition for the contact
curvatures (\ref{eq:boundeqlin3}/\ref{eq:boundeqnonlin2}) (see 
\cite[Sec.\ 12, problem 6]{LaLi_elast}, \cite{Seifert90}, and \cite{wrapping2}).
In Ref.~\cite{wrapping2} a quick derivation can be found for the axisymmetric
case in the constant height ensemble:
varying the point of contact changes the energy of the free profile but also the
energy due to the part at the tip. By setting the total variation to zero one
obtains the well-known contact curvature condition
(Eqn.~(\ref{eq:boundeqlin3}) in Monge gauge). Observe that this assumes
differentiability of the energy as a function of contact point position. In the
force ensemble an extra term $\tilde{F}\delta h$ has to be added to the
variation of the bound membrane. A term that is equal and opposite, however,
enters the variation of the free membrane via the
Hamiltonian~(\ref{eq:Hamiltonian}). In total, both terms cancel and one again
obtains the same condition (Eqn.~(\ref{eq:boundeqnonlin2}) in angle-arclength
parametrization).

The remaining condition~(\ref{eq:boundeqnonlin3}) stems from the fact that 
the total arclength is not a conserved quantity, which it would be if we used a 
fixed interval of integration. Relaxing this unphysical constraint requires the 
Hamiltonian to vanish \cite{Hamiltonian95,CastroVillarealGuven06}.


\section{Calculation of the force via the stress tensor \label{app:diffgeostress}}

If the shape of the free membrane is known, the stress tensor $\VECf^{a}$
($a\in\{1,2\}$) can be evaluated at every point of the surface
$\Sigma_{\text{free}}$. The integral of its flux through an arbitrary contour
$\gamma$ which encloses the tip gives the force \cite{surfacestresstensor}
\begin{equation}
  F = - \VECe_{h} \cdot \oint_{\gamma} \romd s \,
    \Big\{ \Big[ \frac{\kappa}{2}(K_{\perp}^{2} - K_{\|}^{2}) - \sigma \Big]
      \, \VECl
    - \kappa (\nabla_{\perp}K) \, \VECn  \Big\}
  \; .
  \label{eq:forceviastresstensor}
\end{equation}
The normal vectors $\VECl$ and $\VECn$ are perpendicular to $\gamma$ and to each
other in every point of the curve. In addition, $\VECl$ is tangential to
the surface whereas $\VECn$ is normal to it. $K_{\perp}$ and $K_{\|}$ are the
curvatures perpendicular (in direction of $\VECl$) and tangential to the curve.
The symbol $\nabla_{\perp}$ denotes the derivative along $\VECl$.

In angle-arclength parametrization, the curvatures are given by: 
$K_{\perp}=-\dot{\psi}$, $K_{\|}=-\sin{(\psi)}/\rho$, and $K=-p_{\psi}/(2\rho)$. 
Eqn.~(\ref{eq:forceviastresstensor}) can then be written as
\begin{eqnarray}
  F & = & -\frac{\kappa}{2} \oint_{\gamma} \romd s \;
    \Big\{ \Big[ \Big(\dot{\psi}^{2} - \frac{\sin^{2}{\psi}}{\rho^{2}}\Big) 
      - \sigma \Big] \sin{\psi} 
  \nonumber \\
  && \qquad\qquad\quad + \; \frac{1}{\rho} 
    \Big( \dot{p}_{\psi} - \frac{p_{\psi}}{\rho}\dot{\rho} \Big) \cos{\psi}
    \Big\}
  \; .
\end{eqnarray}
The integrand can be evaluated further by inserting the Hamilton 
equations~(\ref{eq:shapeequationsnonlinear}) and making use of the fact that the 
Hamiltonian~(\ref{eq:Hamiltonian}) is zero. One obtains
\begin{equation}
  F = \frac{\kappa}{2} \oint_{\gamma} \romd s \; 
    \Big(\frac{p_{z}+\tilde{F}}{\rho}\Big)
  \; .
\end{equation}
If we now exploit axial symmetry by integrating around a circle of radius 
$\rho=\Rint$, we finally get $F=p_{z}+F$; the momentum $p_{z}$ conjugate to $z$ has 
to vanish identically which implies that the Lagrange multiplier function 
$\lambda_{z}$ is equal to $\tilde{F}$. This at first maybe surprising 
result is no coincidence at all. In fact, in 
Ref.~\cite{CastroVillarealGuven06} it was shown that the Lagrange multiplier 
functions which fix the geometrical constraints are closely related to the 
external forces via the conservation of stresses. 

Expression~(\ref{eq:forceviastresstensor}) can also be translated into ``Monge gauge''. 
If we again exploit axial symmetry and integrate around a circle of radius $\rho=\Rint$, 
it reads
\begin{eqnarray}
  F & = & -2 \pi \Rint
    \Bigg\{ \Big[ \frac{\kappa}{2}
      \Big(\frac{h''(\rho)^{2}}{g^{3}} - \frac{h'(\rho)^{2}}{\rho^{2}g}\Big)
        - \sigma \Big] \frac{h'(\rho)}{\sqrt{g}}
  \nonumber \\
  && \qquad \qquad + \; \kappa \Big(
  \frac{h''(\rho)}{\sqrt{g}^{3}} + \frac{h'(\rho)}{\rho \sqrt{g}} \Big)'
    \; \frac{1}{g} \Bigg\}\Bigg|_{\rho=\Rint}
  \!\! ,
  \label{eq:forceviastresstensorMonge}
\end{eqnarray}
where $g = 1 + h'(\rho)^{2}$. Note that the dash denotes derivatives with
respect to $\rho$.

If in particular we choose to evaluate the force at $\Rint=\Rp$, the
expression~(\ref{eq:forceviastresstensorMonge}) simplifies
considerably to Eqn.~(\ref{eq:forcelinear}).


\section{Numerical calculations \label{app:numericalmethod}}

The Hamilton equations~(\ref{eq:shapeequationsnonlinear}) were solved by using
a shooting method \cite{NumericalRecipes}: for a trial contact point $c$
Eqns.~(\ref{eq:shapeequationsnonlinear}) were integrated with a fourth-order
Runge-Kutta method. The value of $c$ determined the contact angle $\alpha$ and
with it $\psi$, $\rho$, $p_{\psi}$, and $p_{\rho}$ at $s=\underline{s}$ via the
boundary conditions~(\ref{eq:boundaryequationsnonlinear}).
The integration was stopped as soon as $\rho$ was equal or greater than $\Rp$.
To reach $\Rp$ exactly one extra integration with the correct stepsize
backwards was performed.
Finally, the value(s) of $c$ for which $\psi=0$ at $\Rp$ were identified for
fixed parameters $F$, $\sigma$, $w$, etc.

If the calculation had been done in the constant height ensemble, one would
additionally have to check whether the correct indentation $h_{0}$ was reached
at $\rho=\Rp$ after shooting. In the constant force ensemble this complication
of meeting a second condition is avoided which is why we chose to use it for the
nonlinear calculations.




\begin{thebibliography}{99}
\bibitem{substrate}
Note that the part of the system outside the pore just acts like a
reservoir fixing the lateral tension. Its energetics do not have
to be considered explicitly.
\bibitem{signconvention}
The convention used in this article is that a sphere with outward pointing
normal vector has a positive $K$.
\bibitem{Gaussiancurvature}
The Gauss-Bonnet theorem states that:
\begin{equation*}
  \int_{\Sigma} \romd A\; K_{\text{G}}
    = 2\pi - \int_{\partial{\Sigma}} \!\! \romd s\; K_{\text{g}}
  \; ,
\end{equation*}
for a simply connected surface \cite{DifferentialGeometry}. In our case the
boundary $\partial\Sigma$ of the surface $\Sigma$ is a circle of radius $\Rp$.
Its geodesic curvature $K_{\text{g}}$ is equal to $\Rp^{-1}$, such that the
second integral yields $2\pi$. Thus, the integral over the Gaussian curvature
$K_{\text{G}}$ is zero as long as no topological changes occur.
\bibitem{axisintercept}
The axis intercept is needed to gauge the distance between AFM tip and substrate
in the experiment and can therefore not be extracted as a second parameter.
\end{thebibliography}
\end{document}